\documentclass[12pt,aps,nofootinbib]{revtex4}
\usepackage{amsmath}
\usepackage{graphicx}

\def\gsim{\gtrsim}

\newcommand{\Gammamat}{\boldsymbol \Gamma}
\newcommand{\Rmat}{\mathbf R}
\renewcommand{\vec}[1]{\boldsymbol #1}

\DeclareMathOperator{\Trace}{Tr}

\begin{document}

\title{Pairing in Asymmetric Many-Fermion Systems:\\ Functional Renormalisation Group Approach}
\author{Boris Krippa$^{1}$}
\affiliation{$^1$School  of Science and Technology,
Nottingham Trent University, NG1 4BU, UK}
\date{\today}
\begin{abstract}
Functional renormalisation group approach is applied to a imbalanced
many-fermion system with a short-range attractive force. We introduce a composite 
boson field to describe pairing effects, and assume a simple ansatz for 
the effective action. A set of approximate flow equations for 
the effective coupling including boson and fermionic fluctuations is derived and solved. We identify the critical values of particle number density mismatch when the system undergoes to a normal state. We determine the phase diagram both at unitarity and around.  The obtained phase diagram is in a reasonable agreement with the experimental data. 

\end{abstract}
\maketitle

The mechanism of pairing in imbalanced many-fermion systems is nowdays a subject of the intensive theoretical and experimental studies (see ref. \cite{Sto} for review). This phenomena occurs in many physical systems from molecular physics to quark matter at finite density. Being different in details, the underlying dynamical mechanisms share a common feature related to Cooper instability leading to a rearrangement of the ground state and associated spontaneous symmetry breaking.

In this paper we focus on the asymmetric ultracold atomic Fermi mixture of two fermion flavours, which realizes a highly tunable system of strongly interacting fermions. This tunability is provided by a Feshbach resonance, which allows to control the interaction strength between two different species of fermions and explore the BEC-BCS crossover in a wide range of physical parameters. Another tunable parameter (in asymmetric systems) is the population imbalance which can be used to probe how stable the superfluid phase is. The problem was studied long time ago by Clogston and Chandrasekhar \cite{CC}  who found that in the BCS limit the system with the chemical potential mismatch $\delta\mu$ undergoes first order phase transition to a normal phase at $\delta\mu = 0.71 \Delta_0$ where $\Delta_0$ is the gap at zero temperature for balanced system. Recently, the issue has been looked at again but now in the case of strongly interacting fermions with infinite scattering length (unitary limit) \cite{Sto}. Most theoretical studies have been performed in the framework of the mean-field (MF) type of approaches which are of limited use for the imbalanced many-fermion systems and may not be reliable in  providing quantitative answers. In many cases the effects of quantum fluctuations turn out to be important.

The aim of the present paper is to set up a framework to study pairing phenomena in imbalanced many-fermion systems using the formalism of Functional Renormalisation Group \cite {Wet}  (FRG) where the effects of quantum fluctuations are included in a consistent and reliable way. The  FRG approach makes use of  the Legendre transformed  effective action: $\Gamma[\phi_c]=W[J]-J\cdot \phi_c$, where $W$ is the usual partition function in the presence of an external source $J$. The action functional $\Gamma$ generates the 1PI Green's functions and it reduces to the effective potential for homogeneous systems. In the FRG  one introduces an artificial renormalisation group flow, generated by a momentum scale $k$ and we define the
effective action by integrating over components of the 
fields with $q \gsim k$. The RG trajectory then interpolates between the 
classical action of the underlying field theory (at large $k$), and the 
full effective action (at $k=0$). This method has
been successfully applied to a range of problems, from condensed matter
physics \cite{Wet2} to particle physics \cite{Gie}.

The evolution equation for $\Gamma$ in the ERG has a 
one-loop structure  and can be written as
\begin{equation}
\partial_k\Gamma=-\frac{i}{2}\,\Trace \left[
(\Gammamat ^{(2)}_{BB}-\Rmat_B)^{-1}\,\partial_k\Rmat_B\right]
+\frac{i}{2}\,\Trace \left[
(\Gammamat ^{(2)}_{FF}-\Rmat_F)^{-1}\,\partial_k\Rmat_F\right].
\label{eq:Gamevol}
\end{equation}
Here $\Gammamat ^{(2)}_{FF(BB)}$ is the matrix containing second
functional derivatives of the effective action with respect to the
fermion (boson) fields and $\Rmat_{B(F)}$ is a matrix containing the
corresponding boson (fermion) regulators which must vanish when the running scale approaches zero.
  A $2\times 2$ matrix structure 
arises for the bosons because we treat $\phi$ and $\phi^\dagger$ as
independent fields in order to include the number-violating condensate.
A similar structure also appears for the fermions. By inserting the
ansatz for $\Gamma$ into this equation one can turn it into a set of
coupled equations for the various couplings.

Here we study a system of fermions with population imbalancies interacting through an attractive 
two-body point-like potential and consider pairing between the fermions with different flavours assuming that
the interaction between the identical ones is negligible. We take as our starting point an EFT that 
describes the $s$-wave scattering of two nonidentical fermions with a $T$-matrix 
determined by the scattering length $a$.
A positive scattering length corresponds to a system with a two-body 
bound state (and hence repulsive phase-shifts for low-energy scattering) 
whereas a negative scattering length  corresponds to one without a bound state.
The binding energy gets deeper as $a$ gets smaller,
while the limit $a\rightarrow\pm\infty$ is related to a zero-energy 
bound state.

Since we are interested in the appearance of a gap in the fermion
spectrum, we need to parametrise our effective action in a way that
can describe the qualitative change in the physics when this occurs.
A natural way to do this is to introduce a boson field whose vacuum 
expectation value (VEV) describes the gap and so acts as the 
corresponding order parameter.  At  
the start of the RG flow, the boson 
field is not dynamical and is introduced through a 
Hubbard-Stratonovich transformation of the four-fermion pointlike interaction.
As we integrate out more and more of the fermion degrees of freedom by 
running $k$ to lower values, we generate dynamical terms in the bosonic
effective action.

We take the following ansatz for $\Gamma$ which is a generalisation of the ansatz used in \cite{Kri1} for a balanced many-fermion system
\begin{eqnarray}
\Gamma[\psi,\psi^\dagger,\phi,\phi^\dagger,\mu,k]&=&\int d^4x\,
\left[\phi^\dagger(x)\left(Z_\phi\, i \partial_t 
+\frac{Z_m}{2m}\,\nabla^2\right)\phi(x)-U(\phi,\phi^\dagger)\nonumber\right.\\
&&\qquad\qquad+ \sum_{i=1}^{i=2}\psi^\dagger\left( Z_\psi (i \partial_t+\mu_i)
+\frac{Z_{M_i}}{2M_i}\,\nabla^2\right)\psi\nonumber\\
&&\qquad\qquad\left.- g\left(\frac{i}{2}\,\psi^{\rm T}\psi\phi^\dagger
-\frac{i}{2}\,\psi^\dagger\psi^{\dagger{\rm T}}\phi\right)\right],
\label{eq:Gansatz}
\end{eqnarray}
Here $M_i$ and $m$ are masses of fermions and composite boson. All renormalisation factors, couplings and chemical potentials run with the scale $k$. The term containing the boson chemical potential is quadratic in $\phi$ so it can be absorbed into effective potential $U$ and the Yukawa coupling is assumed to describe the decay (creation) of a pair of nonidentical fermions. Due to $U(1)$ symmetry the effective potential depends on the combination $\phi^{\dagger}\phi$. We expand the potential $U(\rho)$ near its minima  and keep terms up to order $\rho^3$.
\begin{equation}
U(\phi,\phi^\dagger)= u_0+ u_1(\rho-\rho_0)
+\frac{1}{2}\, u_2(\rho-\rho_0)^2 + \frac{1}{6}\, u_3(\rho-\rho_0)^3 + . . . ,
\label{eq:potexp}
\end{equation}
where $\rho = \phi^{\dagger}\phi$. We  assume $Z_{\psi_i} = Z_{M_i} = 1$ and neglect running of Yukawa coupling.
One notes that the expansion near minimum of the effective potential (either trivial or nontrivial), being quite reliable in the case of  second order phase transition, may not be
 sufficient to quantitatively describe the first order one. It is worth emphasizing that  the CC limit related transition from the superfluid phase to a normal one is of the first order so 
that a reliability of the expansion needs to be verified.  However, as we will discuss below, at small/moderate asymmetries even a simple ansatz for the effective action  the effective potential expanded up to the third order in the field bilinears gives a reasonable description of the corresponding phase diagram and provides a clear evidences that the phase transition is indeed of first order.
 
At the starting scale the system is in a symmetric regime with a trivial minimum so that $u_{1}(k)$ is positive. At some lower scale $k = k_{crit}$ the coupling $u_{1}(k)$ becomes zero and the system undergoes a transition to the broken phase with a nontrivial minimum and develops the energy gap.

 In our RG evolution we have chosen the trajectory when chemical potentials run in the broken phase and the corresponding particle densities $n_i$ remain fixed so that we define
"running Fermi-momenta" for two fermionic species as $p_i = \sqrt{2 M_i \mu_i}$. It is convenient to work with the total chemical potential and their difference so we define
\begin{equation}
\mu = \frac{\mu_1 + \mu_2}{2}; \qquad \delta = \frac{\mu_1 - \mu_2}{2}
\end{equation}
and assume that $\mu_1$ is always larger then $\mu_2$. Calculating corresponding functional derivatives, taking the trace and performing a contour integration results in the following flow equation for the effective potential
\begin{eqnarray}
\partial_k U
&=&-\,\frac{1}{2 Z_\psi}\int\frac{d^3{\vec q}}{(2\pi)^3}\,\frac{E_{1F} + E_{2F}}
{\sqrt{(E_{1F} + E_{2F})^2+ 4 g^2\rho}}\,(\partial_k
R_{1F} + \partial_k R_{2F})\nonumber\\
\noalign{\vskip 5pt}
&&+\,\frac{1}{2Z_\phi}\int\frac{d^3{\vec q}}{(2\pi)^3}\,
\frac{E_{BR}}{\sqrt{E_{BR}^2-V_B^2}}
\,\partial_kR_B,\label{eq:potevol}
\end{eqnarray}
where
\begin{equation}
E_{BR}(q)=\frac{Z_m}{2m}\,q^2 + U'' \rho + U' + R_B(q,k),\qquad
V_B = U''\rho,
\end{equation}
and 
\begin{equation}
E_{iF} \equiv E_{iF}(q,k,p_i)=\epsilon_{i}(q)-\mu_i+R_{iF}(q,p_i,k),\qquad \epsilon_{i}(q) = q^2/2 M_i.
\end{equation}

Here we denote $U' = \frac{\partial U}{\partial \rho}$ and $U'' = \frac{\partial^2 U}{\partial \rho^2}$ etc.

One notes that the position of the pole in the fermion loop integral which defines the corresponding dispersion relation is given by

\begin{equation}
q_0 = \frac{E_{2F} - E_{1F} \pm \sqrt{(E_{2F} + E_{1F})^2+ 4 \Delta^2}}{2},
\end{equation}

where $\Delta^2 = 4 g^2 \rho$ is the square of the pairing gap.

In the physical limit of vanishing scale this dispersion relation indicates a possibility of the gapless exitation in asymmetric many-fermion systems (much discussed Sarma phase \cite{Sar}). The gappless
 exitation occurs  at $\frac{\Delta}{\delta} <$ 1.  As we will show below,  this condition is never fulfilled so that Sarma phase does not occur.
We note, however, that this conclusion is valid at zero temperature case and can be altered at finite temperature where the possibility for the Sarma phase  still exists\cite{Sto}.
The corresponding bosonic exitations are just gapless "Goldstone" bosons as it should be.

In order to follow the evolution at constant density and running chemical potential we  define the total derivative
\begin{equation}
d_k=\partial_k+(d_k\mu)\,\frac{\partial}{\partial\mu} + (d_k\rho)\,\frac{\partial}{\partial\rho}  , 
\end{equation}
where $d_k\mu=d\mu/dk$,   $d_k\rho=d\rho/dk$. Applying this to effective potential,  
demanding that $n$ is constant ($d_k n=0$) gives
the set of the flow equations 
\begin{eqnarray}
2Z_{\phi}\,d_k\rho - \chi d_k \mu
&=&\left.\frac{\partial}{\partial \mu}
\Bigl(\partial_k \bar U\Bigr)\right|_{\rho=\rho_0},\\
\noalign{\vskip 5pt}
d_k u_0+n\,d_k\mu
&=&\left.\partial_k U\right|_{\rho=\rho_0},\\
\noalign{\vskip 5pt}
-u_2\,d_k\rho + 2Z_\phi\,d_k\mu
&=&\left.\frac{\partial}{\partial \rho}
\Bigl(\partial_k U\Bigr)\right|_{\rho=\rho_0},\\
\noalign{\vskip 5pt}
 d_k u_2 -u_3 d_k \rho - d_k\mu\beta&=&\left.\frac{\partial^2}{\partial \rho^2}
\Bigl(\partial_k U\Bigr)\right|_{\rho=\rho_0},\\
\noalign{\vskip 5pt}
\frac{1}{2}\chi' d_k \mu + d_k Z_\phi + \frac{1}{2}\beta d_k \rho &=&-\,\frac{1}{2}\left.\frac{\partial^2}{\partial \mu\partial\rho}
\Bigl(\partial_k U\Bigr)\right|_{\rho=\rho_0},\\
\noalign{\vskip 5pt}
-\beta' d_k \mu + d_k u_3 - u_4 d_k \rho &=&\left.\frac{\partial^3}{\partial\rho^3}
\Bigl(\partial_k U\Bigr)\right|_{\rho=\rho_0}
\end{eqnarray}

where we have defined
\begin{equation}
\chi = \frac{\partial^{2} U}{\partial \mu^2},  \qquad  \chi' = \frac{\partial^{3} U}{\partial \mu^2 \partial \rho}, 
\qquad  \beta = \frac{\partial^{3} U}{\partial \mu \partial \rho^2}, \qquad  \beta' = \frac{\partial^{4} U}{\partial \mu \partial \rho^3}
\end{equation}

The left-hand sides of these equations contain a number of higher order terms  such as $u_4$, $\chi$, $\chi'$, $\beta$, $\beta'$.  The scale dependence of   these couplings is obtained from evolution with fermion loops only.

The driving terms in these evolution equations are given by appropriate 
derivatives of Eq.~(\ref{eq:potevol}). In the symmetric phase we evaluate 
these expressions at $\rho=0$. The driving term for the chemical potential evolution 
vanishes in this case, and hence $\mu$ remains constant. In the broken phase 
we keep $\rho$ non-zero and set $u_1=0$. 
The details of the derivation can be found in ref. \cite{Kri2}.

Neglecting the effect of bosonic fluctuations leads to the mean-field expression for the effective potential 

\begin{equation}
 U = -\frac{M_r}{2\pi a} g^2\rho + \frac{1}{2}\int\frac{d^3{\vec q}}{(2\pi)^3}\left[\bar{E}_{1F}+ \bar{E}_{2F}  +  \frac{2 g^2\rho}{\epsilon_1 + \epsilon_2} - \sqrt{(\bar{E}_{1F}+ \bar{E}_{2F})^2+ 4 g^2\rho}\right].
\end{equation}

Here $\bar{E}_{iF} = E_{iF}(q,k,0)$, $a$ is the fermion-fermion scattering length and $M_r$ is the reduced mass.
Imposing the condition $\partial_\rho U|_{\rho=\rho_0} = 0$ we recover the BCS-like gap equation

\begin{equation}
-\frac{M_r}{2\pi a} + \int\frac{d^3{\vec q}}{(2\pi)^3}\,\left[\frac{1}{\epsilon_1 + \epsilon_2} - \frac{1}{{\sqrt{(\bar{E}_{1F} + \bar{E}_{2F})^2+ 4 g^2\rho}}}\,\right]= 0,
\end{equation}

Our approach can be applied to any type of many-fermion system but for a concretness we use a parameter set relevant to nuclear matter: $M_1 = M_2 = 4.76$ fm$^{-1}$, $p_{1(2)} \simeq 1$ fm$^{-1}$ and large fermion-fermion scattering length ($a \>> 1$) fm.
We use the regulators in the form suggested in \cite{Lit} for both bosons and fermions

\begin{equation}
R_{F_i} = \frac{1}{2M_i}\left[(k^{2} sgn(q - p_{\mu_i})- (q^2 -  p^{2}_{\mu_i}))\right]\theta(k^2  - \mid q^2  - p^{2}_{\mu_i}\mid),
\end{equation}
where $p_{\mu_i} = (2 M_i \mu_i)^{1/2}$, and
\begin{equation}
R_B = \frac{1}{2m}((\sigma k)^2 - q^2)\theta(k -q),
\end{equation}
where $\sigma$ is a parameter which defines the relative scale of the bosonic and fermionic regulators. We set $\sigma$ = 1 in the following. The initial conditions for $u'$s and $Z$ can be obtained by differentiating the expression for the effective potential at the starting scale $k = k_{st}$ and setting the parameter $\rho$ to zero.

Now we turn to the results.  First we note that the system undergoes the transition to the broken phase at critical scale $k_{cr} \simeq \frac{p_1 + p_2}{2}$. Its value slowly decreases when the asymmetry is increased while keeping the total chemical potential fixed. We found that the value of the $k_{kr}$ is practically insensitive to the starting scale provided the scale is chosen to be larger than 10 fm$^{-1}$.  The position of the minimum of the effective potential in the unitary regime changes rather slowly with increasing $\delta\mu$until the chemical potential mismatch  reaches some critical value $\delta\mu_{crit} = 0.68\mu$. When $\delta\mu$ becomes larger then $\delta\mu_{crit}$ the minimum of the effective potential drops to zero value thus indicating first order phase transition similar to the CC limit, obtained in the limit of weak coupling.

\begin{figure}
\begin{centering}
\includegraphics[width=10cm]{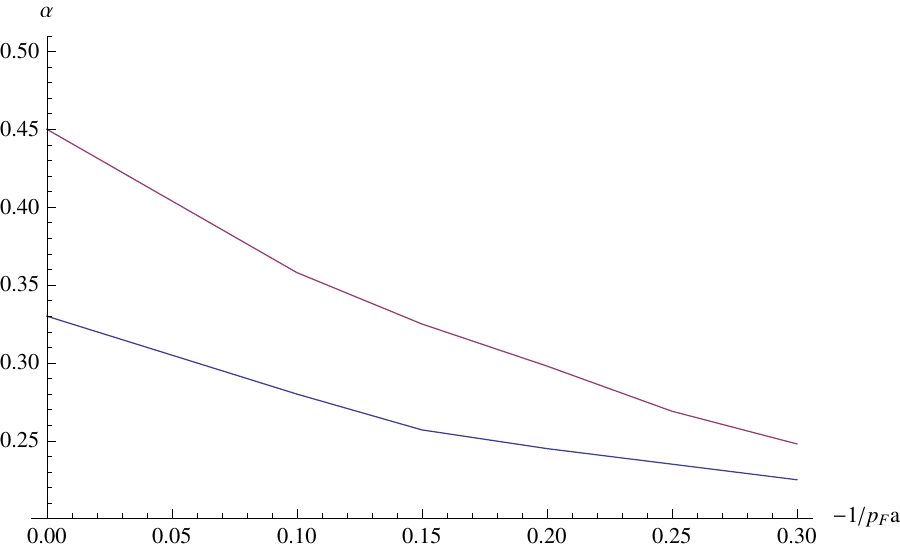}
\par\end{centering}

\caption{Phase diagram as a function of -1/$p_F a$ and polarisation with $p_F$ corresponding to the fermions with a larger density. The upper curve (red online) is the result 
of the calculations and lower curve (blue online) corresponds to experimental data from \cite{Shin}}

\end{figure}

In Fig.1 we show the results for the critical line, separating the gapped and normal phases as a function of the dimensionless parameter $1/p_F a$, where $p_F$ corresponds to the
state with larger density and particle density asymmetry $\alpha = \frac{n_1 - n_2}{n_1 + n_2}$.
 The experimental data are from  \cite{Shin}. The lower curve  is the exponential fit of the data from \cite{Shin}. 
Our theoretical curve approaches  the  fit  with decreasing $p_F |a|$ although always lies  above the experimental data thus indicating the room for a further improvement of the ansatz.  One notes, that at any value of -1/$p_F a$ the phase transition always takes place when $\frac{\Delta}{\delta}$ is greater then one. It means that the  condition required for the Sarma phase is never reached. One can therefore conclude that, at least at $T = 0$, Sarma phase never occurs and consequently the phase  transition is indeed of first order otherwise we would find  that at some point the ratio $\Delta/\delta$ becomes less then one. Physically it means that the system must be viewed as an inhomogeneous mixture of the gapped and normal phases, as suggested in \cite{Bed}\\
The higher order  couplings bring in the corrections on the level of 18-20 $\%$ so the expansion of the effective potential near minimum converges reasonably well. Certainly,  in order to improve the description of the experimental data  the full solution for the unexpanded potential is required but a qualitative conclusion about phase transition being of first order will remain the same regardless of the way the effective potential is treated.

We show on Table 1 the results of the calculations for the superfluid gap in the limit of small density imbalance $\alpha = 0.03$ in comparison with the experimental data from \cite{Shin1}.
 As in the case of the phase diagram the theoretical points are not  far from the experimental data but still lie above them indicating that higher order terms should be included in
 our truncation
 for the effective action to achieve better agreement with the data.\\
\begin{table}[ht]
 \caption{Superfluid gap}
 \centering 
\begin{tabular}{c c c c} \hline\hline
 $1/p_{F}a$ & $\Delta$(exp) & $\Delta$(calc)\\ [0.5ex]	
 \hline 0&0.44&0.55\\
 -0.25&0.22&0.27 \\
  [1ex]
 \hline
 \end{tabular} 
\label{table:nonlin}
 \end{table}

 We have also calculated the critical value of the chemical potential mismatch $\delta\mu_c$ with parameters typical for neutron matter (scattering length $a_{nn} \simeq -18.6$) fm. Again at large
 enough $\delta\mu > \delta\mu_{crit}$ the pairing is disrupted and the system undergoes to a normal phase. The value of  $\delta\mu_c$ can be important for the phenomenology of neutron stars
 because the
 transport properties of the  normal and superconducting phase are very different \cite{Gez1}. Our calculations gives the value  $\delta\mu_c = 0.33
\mu$ to be compared with  the QMC based
 results $\delta\mu_c = 0.27\mu_c$ \cite{Gez2}.

As we mentioned in the introduction the other possible type of imbalance can be caused by the fermion mass mismatch. Our approach is general enough to incorporate this case without any changes of the formalism. In this paper we consider a special case when the Fermi-momenta of two fermionic species are equal thus ruling out the LOFF phase. The general case of the combined mass/density asymmetry with the LOFF phase taken into account will be reported elsewhere.

\begin{figure}
\begin{centering}
\includegraphics[width=10cm]{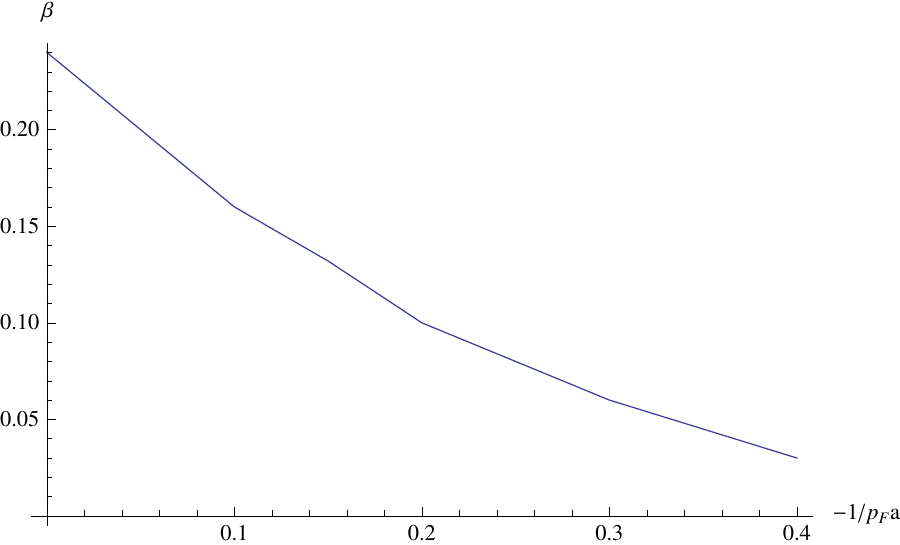}
\par\end{centering}

\caption{Phase diagram as a function of the mass imbalance $\beta$ and  -1/$p_F a$ with $p_F$ corresponding to the fermions with a larger density}. 

\end{figure}

Our result is shown in Fig.2 in the form of the phase diagram as a function of the relative mass imbalance defined as $\beta = \frac{M_1 - M_2}{M_1 + M_2}$ and the dimensionless parameter $1/p_F a$. Without a loss of generality we  assume that $M_1$ is greater then $M_2$. The shape of the curve is  similar to that for the case of the density imbalance although numerically the system in the gapped regime tolerates rather smaller values of mass imbalance compared to case of unequal  densities. Again, the area under the curve corresponds to the gapped phase and one above the curve is the phase with the unpaired fermions.

One notes that we do not consider the BEC region of the phase diagram although the formalism allows to do that. The reason is that in the BEC regime (unlike the unitary and BCS regimes) the results show a sensitivity to the parameter $\sigma$  of the boson cutoff function. Similar situation was found in \cite{Kri4} for the process of low-energy dimer-dimer scattering. Varying $\sigma$ around  the so called optimal choice  \cite{Paw} one could bring the calculation to a reasonable agreement with the experimental data but 
it should be interpreted as the fit, rather than the theoretical prediction.
This sensitivity signals that one needs to include higher order terms in our ansatz for the effective action in order to achieve a better stability and reliably describe the BEC part of the phase diagram.

 In general, one can conclude that, in spite of  a relative simplicity of the assumed ansatz for the effective action  FRG provides a good starting point for a  reasonable description of the phase diagram of asymmetric many-fermion systems. The phase transition is found to be of first order in agreement with the other theoretical results \cite{Pit} and the Sarma phase never occur for this system (at zero temperature) which means that the system should be interpreted as an inhomogeneous mixture of the gapped and normal phases. 

One of the most obvious improvements of our approximation is to use a complete effective potential instead of expanding it near a scale dependent minimum. However, it is very likely that the higher order  terms will result in moderate corrections thus leaving the qualitative conclusion unchanged.

Another potentially important improvement of the formalism would be an  inclusion of the 
fermion-fermion interaction in the particle-hole (ph) channel leading to the Gorkov-Melik-Barkhudarov (GMB) corrections \cite{Gor}. The FRG based studies of the GMB corrections have been performed in \cite{Die} for the case of the balanced many-fermion systems. A generalisation of the approach developed in \cite{Die} to the imbalanced systems is highly nontrivial and requires a serious technical and computational effords, Although the full size FRG calculations including the ph channel are beyond the scope of this paper some preliminary results  indicate that the inclusion of the particle-hole interactions brings the theoretical results closer to the experimental data \cite{Kri3} and, being extended to finite temperature, may significantly alter the position of the corresponding critical point.

\section{acknowledgement}
The author is grateful to M. Birse and N. Walet for valuable discussions.


\begin{thebibliography}{16}

\bibitem{Sto}K. W. Gubbels, H. T. C. Stoof, arXiv:1205.0568 (cond-mat)

\bibitem{CC}B. Chandrasekhar, Appl. Phys. Lett. B\textbf{1}, 7 (1962);  A. Clogston, Phys. Rev. Lett. B\textbf{9}, 266 (1962) 

\bibitem{Wet}C. Wetterich, Phys. Lett. B\textbf{301}, 90 (1993),
T. Morris, Phys. Lett. B\textbf{334}, 355 (1994) {[}arXiv:hep-ph/9403340{]}, D.-U. Jungnickel and C. Wetterich, Phys. Rev. D\textbf{53},
5142 (1996) {[}arXiv:hep-ph/9505267{]}. 

\bibitem{Wet2} J. Berges, N. Tetradis and C. Wetterich, Phys. Rept.
\textbf{363}, 223 (2002) {[}arXiv:hep-ph/0005122{]}, B. Delamotte,
D. Mouhanna and M. Tissier, Phys. Rev. B\textbf{69}, 134413 (2004)
{[}arXiv:cond-mat/0309101{]}. 

\bibitem{Gie}H. Gies, arXiv:hep-ph/0611146{]}. 

\bibitem{Kri1} M. C. Birse, B. Krippa, J. A. McGovern and N. R. Walet,
Phys. Lett. B\textbf{605}, 287 (2005) {[}arXiv:hep-ph/0406249{]}. 

\bibitem{Sar}G. Sarma, J. Phys. Chem. Solids \textbf{24},
1029 (1963). 

\bibitem{Kri2}  B. Krippa, 
J. Phys. \textbf{A39}, 8075 (2006) {[}arXiv:nucl-ph/051283{]}. 

\bibitem{Lit}D. Litim, Phys. Lett.
B \textbf{486}, 92  (2000). 

\bibitem{Shin}Y. Shin et al, 
arXiv:cond-mat/0805.0623, (2008) 

\bibitem{Bed} P. F. Bedaque, H. Caldas and G. Rupak, Phys.Rev.Lett. \textbf{91}, 247002 (2003).

\bibitem{Shin1}A. Schirotzek et al., arXiv:cond-mat.other/0808.0026

\bibitem{Gez1}V. Girigliano, S. Reddy, and R. Sharma,  Phys. Rev.  \textbf{C84}, 045809 (2011). 

\bibitem{Gez2} A. Gezerlis and R. Sharma, Phys. Rev.\textbf{C85}, 015806 (2012).
 
\bibitem{Kri4} M. C. Birse, B. Krippa and N. R. Walet, Phys. Rev.\textbf{A83}, 023621 (2011).

\bibitem{Paw} J. Pawlowski, Annals Phys.\textbf {2831}, 322 (2007).

\bibitem{Pit} S. Giorgini, L. P. Pitaevskii and S. Stringari, Rev. Mod. Phys. \textbf{80}, 1215 (2008).

\bibitem{Gor}L. P. Gorkov and T. K. Melik-Barkhudarov, Sov. Phys. JETP \textbf{13}, 1018 (1961).

\bibitem{Die}S. Floerchinger, M. Scherer, S. Diehl, and C. Wetterich, 
Phys.Rev.\textbf{A78}, 174528 (2008).

\bibitem{Kri3} B. Krippa, work in progress.




\end{thebibliography}
\end{document}